\documentclass[12pt]{article}
\usepackage{amsmath,amssymb,bm,graphicx}
\usepackage{color} 

\begin{document}

\begin{flushright}
\end{flushright}

\vskip 0.5 truecm

\begin{center}
{\Large{\bf No anomalous canonical commutators induced by Berry's phase
}}
\end{center}
\vskip .5 truecm
\centerline{\bf Shinichi Deguchi~$^1$
 {\rm and} Kazuo Fujikawa~$^2$~\footnotetext[2]{Corresponding author.}~\footnotetext[2]{E-mail address: k-fujikawa@riken.jp} }
\vskip .4 truecm
\centerline {\it $^1$~Institute of Quantum Science, College of 
Science and Technology,}
\centerline {\it Nihon University, Chiyoda-ku, Tokyo 101-8308, 
Japan}
\vskip 0.4 truecm
\centerline {\it $^2$~Interdisciplinary Theoretical and Mathematical Sciences Program, 
}
\centerline {\it   RIKEN, Wako 351-0198, 
Japan}

\vskip 0.5 truecm

\makeatletter
\makeatother

\begin{abstract}
The monopole-like singularity of Berry's adiabatic phase in momentum space and associated anomalous Poisson brackets have been recently discussed in various fields. 
With the help of the results of an exactly solvable version of  Berry's model, we show that Berry's phase does not lead to the deformation of the principle of quantum mechanics in the sense of anomalous canonical commutators. If one should assume Berry's phase of genuine Dirac monopole-type, which is assumed to hold not only in the adiabatic limit but also in the non-adiabatic limit, the deformation of  the principle of quantum mechanics could take place.  But Berry's phase of the genuine Dirac monopole-type  is not supported by the exactly solvable version of Berry's model nor by a generic model of Berry's phase. 
Besides,  the monopole-like Berry's phase  in momentum space has a magnetic charge $e_{M}=2\pi\hbar$, for which the possible anomalous term in the canonical commutator $[x_{k},x_{l}]=i\hbar\Omega_{kl}$ would become of the order $O(\hbar^{2})$.
\end{abstract}

\section{Introduction}
It is well-known that the level crossing phenomenon combined with adiabatic approximation in quantum mechanics leads to a monopole-like topological singularity which is commonly called Berry's phase~\cite{Higgins, Berry, Simon}; the level crossing point is a singular point in the adiabatic theorem since the adiabatic theorem in a precise sense states that no level crossing takes place. The monopole-like object is defined in the precise adiabatic limit and thus its behavior away from the precise adiabatic limit, that is required in the analysis of equal-time commutators, for example, is not obvious by looking at only the result of the adiabatic approximation.

We have recently presented an analysis of topological properties of an exactly solvable version of Berry's original model \cite{fujikawa-ap2007,Deguchi-Fujikawa-2019}. We here recapitulate the essence of the analysis which will clearly illustrate what Berry's phase is in the non-adiabatic domain as well as in the adiabatic domain. We consider a magnetic moment placed in a rotating magnetic field $\vec{B}(t)$ which is the original model of Berry~\cite{Berry},
\begin{eqnarray}\label{Schroedinger}
i\hbar\partial_{t}\psi(t)=-\mu\hbar\vec{B}(t)\cdot\vec{\sigma}\psi(t)
\end{eqnarray}
but we choose a specific $\vec{B}(t)$ parameterized by $\varphi(t)=\omega t$ with constant $\omega$, and constant $B$ and $\theta$ with $\vec{\sigma}$ standing for Pauli matrices,
\begin{eqnarray}\label{magnetic field}
\vec{B}(t)=B(\sin\theta\cos\varphi(t), 
\sin\theta\sin\varphi(t),\cos\theta ).
\end{eqnarray}
The exact solution of the Schr\"{o}dinger 
equation \eqref{Schroedinger} is then written 
in the form~\cite{fujikawa-ap2007},
\begin{eqnarray}\label{eq-exactamplitude1}
\psi_{\pm}(t)
&=&w_{\pm}(t)\exp\left[-\frac{i}{\hbar}\int_{0}^{t}dt
w_{\pm}^{\dagger}(t)\hat{H}w_{\pm}(t)\right]\exp\left[-\frac{i}{\hbar}\int_{0}^{t}
\vec{{\cal A}}_{\pm}(\vec{B})\cdot\frac{d\vec{B}}{d t}dt\right]
\end{eqnarray}
with wave functions 
\begin{eqnarray}\label{exact eigenfuntion}
w_{+}(t)&=&\left(\begin{array}{c}
            \cos\frac{1}{2}\Theta(\theta,\eta) e^{-i\varphi(t)}\\
            \sin\frac{1}{2}\Theta(\theta,\eta)
            \end{array}\right), \ \ \ 
w_{-}(t)=\left(\begin{array}{c}
            \sin\frac{1}{2}\Theta(\theta,\eta) e^{-i\varphi(t)}\\
            -\cos\frac{1}{2}\Theta(\theta,\eta)
            \end{array}\right)
\end{eqnarray}
and Berry's phase
\begin{eqnarray}\label{Berry's phase}
\vec{{\cal A}}_{\pm}(\vec{B})\equiv w_{\pm}^{\dagger}(t)(-i\hbar\frac{\partial}{\partial \vec{B}})w_{\pm}(t).
\end{eqnarray}
The variable
\begin{eqnarray}
\Theta(\theta,\eta)=\theta-\alpha(\theta,\eta)
\end{eqnarray} 
is expressed in terms of $\alpha(\theta,\eta)$ which is defined by 
\begin{eqnarray}\label{parameter}
\tan\alpha(\theta,\eta)=\frac{\sin\theta}{\eta + \cos\theta}
\end{eqnarray}
with the parameter
\begin{eqnarray}\label{parameter2}
\eta=\frac{2\mu\hbar B}{\hbar\omega}=\frac{\mu BT}{\pi}
\end{eqnarray}
when one defines the period $T=2\pi/\omega$. The parameter $\eta$ is a ratio of two independent energy scales, the static magnetic energy $\mu\hbar B$ and the kinematical rotational energy $\hbar\omega/2$ of the magnetic field. The dominance of the static energy $\eta\gg 1$ corresponds to the adiabatic limit while the dominance of the kinematical energy $\eta\ll 1$ corresponds to the non-adiabatic limit.

From \eqref{Berry's phase}, by choosing the solution $\psi_{+}(t)$ from now on, we have the static and azimuthally symmetric monopole-like potential with $e_{M}=2\pi\hbar$ \cite{Deguchi-Fujikawa-2019}
\begin{eqnarray}\label{potential1}
{\cal A}_{\varphi} 
= \frac{e_{M} }{4\pi B\sin\theta} \left(1 - \cos\Theta(\theta,\eta) \right)
\end{eqnarray}
and ${\cal A}_{\theta} ={\cal A}_{B} =0$.  
The magnetic charge $e_{M}=2\pi\hbar$ in \eqref{potential1} shows that Berry's phase is an order $O(\hbar)$ quantum effect in the present context. Berry-type phase itself is known to be defined for a non-quantum mechanical process also \cite{Pancharatnam}.  
 
In Fig.1, we show the relation between $\theta$ and $\Theta(\theta,\eta)$~\cite{Deguchi-Fujikawa-2019}.
\begin{figure}[htb]
 \begin{center}
   \includegraphics*[width = 8.0cm]{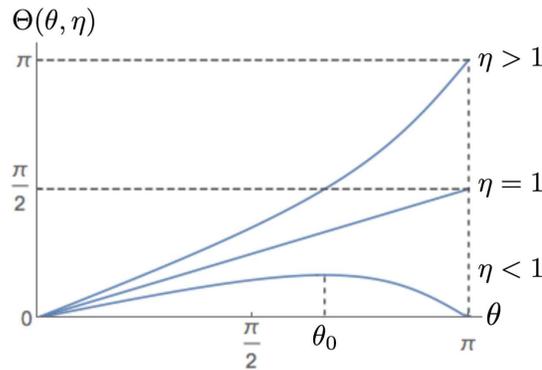}
   \caption{The relation between $\theta$
and $\Theta(\theta,\eta)$ parameterized by $\eta$. Note that $\cos\theta_{0}=-\eta$. See Fig.3 in \cite{Deguchi-Fujikawa-2019}.}\label{Fig_01}
 \end{center}
\end{figure}
The Dirac string \cite{Dirac} which corresponds to the singularity of the potential \eqref{potential1} can appear at $\theta=0$ or $\theta=\pi$; from Fig.1, we see that no singularity at $\theta=0$ since $\Theta(0,\eta)=0$, and the possible Dirac string appears at $\theta=\pi$ for $\Theta(\pi,\eta)=\pi$ (for $\eta>1$) or $\Theta(\pi,\eta)=\pi/2$ (for $\eta=1$) but no string for $\Theta(\pi,\eta)=0$ (for $\eta<1$). The parameter $\eta$ thus controls the topology; we have a monopole-like topology for $\eta>1$ with $\Theta(\pi,\eta)=\pi$ and a Dirac string, while we have a dipole topology for $\eta<1$ with $\Theta(\pi,\eta)=0$ and no Dirac string~\cite{Deguchi-Fujikawa-2019}. 

The magnetic monopole-like configuration is schematically shown in Fig.2 \cite{Deguchi-Fujikawa-2019}.
\begin{figure}[htb]
 \begin{center}
   \includegraphics*[width = 6.0cm]
   {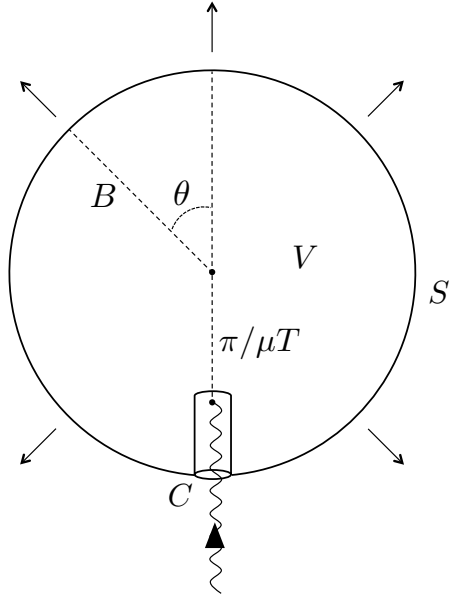}
   \caption{A schematic view of the monopole-like potential \eqref{potential1} for a virtual sphere with radius $B$. The volume V avoids a thin tube surrounding the Dirac string indicated by a wavy line. Geometrical center and the origin of the Dirac string are displaced by the distance $B=\pi/\mu T$. The figure corresponds to $B>\pi/\mu T$, namely, $\eta>1$. Note that the dot at the geometrical center does not represent a singularity, since the monopole-like potential is regular and vanishes at the origin of coordinates. See Fig.5 in \cite{Deguchi-Fujikawa-2019}.}\label{Fig_02}
 \end{center}
\end{figure}
Note that there is no singularity inside the volume $V$ which excludes a thin tube covering the Dirac string indicated by a wavy line in Fig.2. The monopole-like potential is regular and vanishes at the origin of coordinates.
The Dirac string originates at $z=-\pi/\mu T$
on the negative z-axis
corresponding to $\eta=\mu TB/\pi=1$ with fixed $T$. When one approaches the monopole position, which is determined to be the geometrical center on the basis of the observed flux at far away points,  by decreasing $B$ with fixed $T$ in Fig.2, the Dirac string disappears and the dipole topology is realized and, eventually, the monopole-like potential induced by Berry's phase vanishes at the origin $B=0$. This topology change from a monopole to a  dipole is shown to be smooth for {\em fixed} $T$  using the idea of the Dirac string \cite{Deguchi-Fujikawa-2019}, which shows that the naive statement such as "Berry's phase is topologically protected" is not accurate. The monopole-like topology near the origin is a result  of the precise adiabatic  approximation, but the adiabatic approximation fails near the origin in practical applications and thus the monopole topology changes to a dipole topology; the precise adiabatic condition $\eta=\mu TB/\pi \gg 1$ is not maintained for $B=0$ for any finite $T$.

To be more precise, one recovers the conventional Dirac monopole  \cite{Dirac} in the parameter space from \eqref{potential1}, 
\begin{eqnarray}\label{genuine monopole}
{\cal A}_{\varphi}=\frac{e_{M}}{4\pi B\sin\theta}(1-\cos\theta)
\end{eqnarray}  
in the precise adiabatic limit $\eta\rightarrow \infty$, namely, by first letting $T\rightarrow \infty$ for any fixed finite $B$ \cite{Simon}, for which the gap between the geometrical center and the origin of the Dirac string disappears in Fig.2 and $\Theta \rightarrow \theta$ from \eqref{parameter}. One can confirm that \eqref{genuine monopole}
gives the conventional magnetic flux for $\theta\neq \pi$ and $B\neq 0$
\begin{eqnarray}\label{conventional full flux}
{\cal \nabla}\times {\cal A}=\frac{e_{M}}{4\pi B^{2}}{\bf e}_{B}
\end{eqnarray}
with ${\bf e}_{B}=\vec{B}/B$. The equivalent configuration with $\eta\rightarrow \infty$ is realized when observed at $B\rightarrow \infty$ for a fixed nonzero $T$.

On the other hand, in the non-adiabatic limit $\eta\rightarrow 0$ (for example, $B\rightarrow 0$ for any fixed finite $T$, or $T\rightarrow 0$ for any fixed finite $B$),  we have $\Theta(\theta, \eta)\rightarrow 0$ and the vanishing Berry's phase
\begin{eqnarray}\label{trivial phase}
{\cal A}_{\varphi}=0.
\end{eqnarray}
To be more precise, we have a practically useful relation in the non-adiabatic domain $\eta \ll 1$ with $B\neq 0$ \cite{Deguchi-Fujikawa-2019},
\begin{eqnarray}\label{non-adiabatic form}
{\cal A}_{\varphi}\simeq \frac{e_{M}}{8\pi B\sin\theta}(\mu TB\sin\theta/\pi)^{2} 
= \frac{e_{M}}{8\pi }(2\mu /\omega)^{2}B\sin\theta
\end{eqnarray}  
that vanishes quadratically in $T$ for $T\rightarrow 0$ (or $\omega\rightarrow \infty$) with any fixed $B$.
It is also interesting that we have an exact potential (a half-monopole with the magnetic charge $e_{M}/2$ defined by the outgoing flux) at the parameter value $\eta=1$ between the adiabatic and nonadiabatic domains \cite{Deguchi-Fujikawa-2019}
\begin{eqnarray}\label{half sphere}
{\cal A}_{\varphi}&=&\frac{e_{M}}{4\pi B\sin\theta}(1-\cos\frac{1}{2}\theta).
\end{eqnarray}  
But the Dirac string becomes observable in this case, and thus the monopole and the Dirac string combined together become topologically a dipole \cite{Deguchi-Fujikawa-2019}.
The topological structure of Berry's phase is thus remarkably rich.

\section{Berry's phase in a generic level crossing model}

To discuss Berry's phase in momentum space, which is relevant to the analysis of canonical commutators, we recapitulate the analysis of Berry's phase in a generic level crossing problem.  We shall show that Berry's phase at the two limiting cases, at adiabatic and non-adiabatic limits, is accurately described by the exactly solvable model discussed in Section 1.   
Berry's phase is generally described by an effective  Hamiltonian near the  level crossing point
\begin{eqnarray}\label{level crossing Hamiltonian}
H=-\hbar\mu \vec{p}(t)\cdot \vec{\sigma}
\end{eqnarray}
where $\vec{\sigma}$ stands for the pseudo spin that describes the upper and lower crossing levels which are characterized by a generic variable $\vec{p}(t)$, namely, general time dependent $\vec{p}(t)$, such as the magnetic field $\vec{B}(t)$ in the original model of Berry \cite{Berry} with a suitable coupling constant $-\hbar\mu$. We choose this specific notation of the coupling constant to compare the present analysis with the analysis of the original Berry's model \eqref{Schroedinger}; only the combination $\hbar\mu$ has a physical meaning and $\hbar$ here has no quantum mechanical significance we discuss later.  The variable $\vec{p}(t)$ corresponds to the Bloch momentum in the context of condensed matter physics \cite{Karplus, Fang, Hirsch, Niu}. 

For the present purpose to understand the appearance of a monopole-like object in the adiabatic limit of the movement of $\vec{p}(t)$
as well as the smooth disappearance of the monopole-like object in the non-adiabatic limit of the movement of $\vec{p}(t)$, it is important to formulate the problem in a more general manner than the common adiabatic approximation in the Schr\"{o}dinger equation.  The formulation of Berry's phase with a second quantized formalism or equivalent formalism is suited for this purpose \cite{Stone, deguchi, fujikawa-ap2007}.
To analyze the above Hamiltonian, we thus follow the procedure adopted by Stone~\cite{Stone} who initiated the general formulation of Berry's phase, which can be applied to the field theoretical considerations also. 

We thus start with  the examination of the Schr\"{o}dinger equation
\begin{eqnarray}\label{Schroedinger equation}
i\hbar \partial_{t}\psi(t)=H\psi(t)
\end{eqnarray}
 or the Lagrangian  given by
\begin{eqnarray}\label{Lagrangian}
L=\psi^{\dagger}(t)[i\hbar \partial_{t}-H]\psi(t)
\end{eqnarray}
where the two-component spinor $\psi(t)$ specifies the movement of upper and lower levels which appear in the level crossing problem.
We then perform
a time-dependent unitary transformation
\begin{eqnarray}\label{unitary1}
\psi(t)= U(\vec{p}(t))\psi^{\prime}(t),\ \ 
\psi^{\dagger}(t)={\psi^{\prime}}^{\dagger}(t) 
U^{\dagger}(\vec{p}(t))
\end{eqnarray}
with
\begin{eqnarray}
U(\vec{p}(t))^{\dagger}\mu\vec{p}(t)\cdot\vec{\sigma}
U(\vec{p}(t))
=\mu|\vec{p}|\sigma_{3}.
\end{eqnarray} 
This unitary transformation is explicitly given by a $2\times2$ matrix
$U(\vec{p}(t))=\left(
             v_{+}(\vec{p})\  v_{-}(\vec{p}) \right)$,
where
\begin{eqnarray}
v_{+}(\vec{p})=\left(\begin{array}{c}
            \cos\frac{\theta}{2}e^{-i\varphi}\\
            \sin\frac{\theta}{2}
            \end{array}\right), \ \ \ 
v_{-}(\vec{p})=\left(\begin{array}{c}
            \sin\frac{\theta}{2}e^{-i\varphi}\\
            -\cos\frac{\theta}{2}
            \end{array}\right)
\end{eqnarray}
which corresponds to a use of instantaneous eigenfunctions of the operator $\mu\vec {p}(t)\cdot\vec{\sigma}$ where $\vec {p}(t)=|\vec{p}(t)|(\sin\theta\cos\varphi, \sin\theta\sin\varphi, \cos\theta)$ with time dependent $\theta(t)$ and $\varphi(t)$.

Based on this transformation, 
 the equivalence of two 
Lagrangians is derived, namely, $L$ in \eqref{Lagrangian}
and 
\begin{eqnarray}
L^{\prime}=
{\psi^{\prime}}^{\dagger}[i\hbar\partial_{t}
+\hbar\mu|\vec{p}(t)|\sigma_{3}+U(\vec{p}(t))^{\dagger}
i\hbar\partial_{t}U(\vec{p}(t))]\psi^{\prime}.
\end{eqnarray}
 The starting Hamiltonian \eqref{level crossing Hamiltonian} is thus replaced by (using the argument of path integral \cite{Stone})
\begin{eqnarray}\label{geometric-phase}
H^{\prime}(t)&=&
-\hbar\mu|\vec{p}(t)|\sigma_{3}+ U(\vec{p}(t))^{\dagger}
\frac{\hbar}{i}\partial_{t}U(\vec{p}(t))\nonumber\\
&=& -\hbar\mu|\vec{p}(t)|\sigma_{3} -\hbar\left(\begin{array}{cc}
\frac{(1+\cos\theta)\dot{\varphi}}{2}&\frac{\dot{\varphi}\sin\theta+i\dot{\theta}}{2}\\
            \frac{\dot{\varphi}\sin\theta-i\dot{\theta}}{2}&
\frac{(1-\cos\theta)\dot{\varphi}}{2}
            \end{array}\right).
\end{eqnarray}
In the adiabatic limit
\begin{eqnarray}
\mu|\vec{p}(t)| T\gg 2\pi\hbar,
\end{eqnarray}
we have 
\begin{eqnarray}\label{adiabatic Stone phase}
H^{\prime}_{ad} \simeq  -\hbar\mu|\vec{p}(t)|\sigma_{3} -\hbar\left(\begin{array}{cc}
\frac{(1+\cos\theta)\dot{\varphi}}{2}&0\\
            0&
\frac{(1-\cos\theta)\dot{\varphi}}{2}
            \end{array}\right).
\end{eqnarray}
Here $T$ is the period of the dynamical
variable $\vec{p}(t)$ and $2\pi\hbar$ stands for the magnitude of 
the geometric term times $T$, namely,  we estimate $\dot{\varphi} \sim 2\pi/T$.
If $T$ is sufficiently large $\hbar\mu|\vec{p}(t)| T\gg 2\pi\hbar$,
one may neglect the off-diagonal parts in \eqref{geometric-phase} and retain only the diagonal components \eqref{adiabatic Stone phase}.
Stone then finds
 that the adiabatic Berry's phase for the $++$ component~\cite{Stone}
\begin{eqnarray}\label{adiabatic Stone phase2}
&&\exp[-i/\hbar \oint {H^{\prime}}^{(++)}_{ad}dt]\nonumber\\
&=&\exp[i\mu\oint|\vec{p}(t)|dt+i \oint \frac{(1+\cos\theta)}{2}d\varphi],
\end{eqnarray}
namely, the monopole-like flux 
\begin{eqnarray}
\Omega_{mono}=-\hbar\oint \frac{(1+\cos\theta)}{2}d\varphi
\end{eqnarray}
or the monopole-like potential 
\begin{eqnarray}\label{adiabatic Stone phase3}
{\cal A}_{\varphi}=-\hbar\frac{(1+\cos\theta)}{2|\vec{p}|\sin\theta} =\frac{e_{M}}{4\pi|\vec{p}|\sin\theta}(-1-\cos\theta)
\end{eqnarray}
  in the {\em lower hemisphere}
is generated by a (formal) singularity located at the origin of the parameter space $\mu\vec{p}$, where two levels cross. We note that  the magnetic charge of the monopole-like potential \eqref{adiabatic Stone phase3} is given by 
\begin{eqnarray}\label{monopole charge}
e_{M}=2\pi\hbar
\end{eqnarray}
which shows that Berry's phase in the present context is a quantum effect.
From the exact expression \eqref{geometric-phase}, one also sees that the monopole-like potential is a result of adiabatic  approximation. 

It is confirmed using the relation \eqref{geometric-phase} 
that if $\hbar$ times the frequency of $\vec{p}(t)$, $2\pi\hbar/T$, is much larger than the magnitude of the level crossing interaction  $\hbar\mu|\vec{p}(t)|$ or if the particle approaches the monopole position $|\vec{p}(t)|=0$ for any finite $T$, 
\begin{eqnarray}\label{non-adiabatic condition}
\hbar\mu|\vec{p}(t)| T\ll 2\pi\hbar,
\end{eqnarray}
then the geometric term dominates the $\hbar\mu|\vec{p}(t)|\sigma_{3}$ term. 
To see the implications of the non-adiabatic condition \eqref{non-adiabatic condition} explicitly, one may perform a further unitary transformation of the 
fermionic  variable
\begin{eqnarray}
\psi^{\prime}(t)= U(\theta(t))\psi^{\prime\prime}(t), \ \ \  
{\psi^{\prime}(t)}^{\dagger}=
{\psi^{\prime\prime}}^{\dagger}(t) 
U^{\dagger}(\theta(t))
\end{eqnarray}
with~\cite{deguchi}
\begin{eqnarray}\label{unitary2}
U(\theta(t))=\left(\begin{array}{cc}
            \cos\frac{\theta}{2}&-\sin\frac{\theta}{2}\\
            \sin\frac{\theta}{2}&\cos\frac{\theta}{2}
            \end{array}\right)
\end{eqnarray}
in addition to \eqref{unitary1}, which diagonalizes the dominant Berry's phase term.
The Hamiltonian \eqref{geometric-phase} then becomes
\begin{eqnarray}\label{exact non-adiabatic}
H^{\prime\prime}(t)
&=&
-\hbar\mu|\vec{p}(t)| U(\theta(t))^{\dagger}
\sigma_{3}U(\theta(t))\nonumber\\
&&+ (U(\vec{p}(t))U(\theta(t)))^{\dagger}
\frac{\hbar}{i}\partial_{t}(U(\vec{p}(t))U(\theta(t)))
\nonumber\\
&=&
-\hbar\mu|\vec{p}(t)| \left(\begin{array}{cc}
            \cos\theta&-\sin\theta\\
            -\sin\theta&-\cos\theta
            \end{array}\right)
-\hbar\left(\begin{array}{cc}
            \dot{\varphi}&0\\
            0&0
            \end{array}\right).
\end{eqnarray}
Note that the first term is bounded by $\hbar\mu|\vec{p}(t)|$ and the second term is dominant for the non-adiabatic case $\hbar\mu|\vec{p}(t)| T\ll 2\pi\hbar$. We emphasize that both \eqref{geometric-phase} and \eqref{exact non-adiabatic} are exact expressions.  

The Hamiltonian in the non-adiabatic approximation by ignoring small off-diagonal terms then becomes
\begin{eqnarray}\label{non-adiabatic Stone phase}
H^{\prime\prime}_{\rm nonad} \simeq
-\hbar\mu|\vec{p}(t)| \left(\begin{array}{cc}
            \cos\theta&0\\
            0&-\cos\theta
            \end{array}\right)
-\hbar\left(\begin{array}{cc}
            \dot{\varphi}&0\\
            0&0
            \end{array}\right).
\end{eqnarray}
The topological Berry's phase thus either vanishes or becomes trivial in the non-adiabatic limit 
\begin{eqnarray}\label{trivial phase2}
\exp\{-\frac{i}{\hbar}\oint {\cal A}_{\varphi}|\vec{p}|\sin\theta \dot{\varphi}dt\}=\exp\{i\oint \dot{\varphi}dt\}=\exp\{2i\pi\}=1
\end{eqnarray}
independently of $\theta$ for the very rapid movement of $\vec{p}(t)$, namely, $T\rightarrow 0$ with fixed $\hbar\mu|\vec{p}(t)|$, or very close to the monopole position, namely, $\hbar\mu|\vec{p}(t)|\rightarrow 0$ with fixed $T$.
Note that the trivial phase \eqref{trivial phase2} is equivalent to the vanishing Berry's curvature. The regular transformation \eqref{unitary2} may be regarded as a resolution of monopole singularity in Berry's phase by noting that Hamiltonian \eqref{level crossing Hamiltonian} is regular in the variable $\vec {p}(t)$. 

We have thus shown that the monopole-like object associated with generic Berry's phase are parameterized by 
\begin{eqnarray}\label{parameter eta}
\eta=\frac{\mu|\vec{p}|T}{\pi},
\end{eqnarray}
namely, 
\begin{eqnarray}\label{Berry connection}
{\cal A}_{k}(\vec{p}; \eta)\dot{p}_{k} = {\cal A}_{k}(\vec{p}; \mu|\vec{p}|T/\pi)\dot{p}_{k} = {\cal A}_{k}(\vec{p}; 2\hbar\mu|\vec{p}|/\hbar\omega)\dot{p}_{k}.
\end{eqnarray}
The parameter $T=2\pi/\omega$ stands for the {\em typical time scale} of the movement of the variable $\vec{p}(t)$. In the adiabatic limit $\eta\rightarrow\infty$, we have the monopolelike
phase as in \eqref{adiabatic Stone phase3}, while Berry's phase becomes trivial
(i.e., curvature vanishes) as in \eqref{trivial phase2} in the non-adiabatic limit $\eta\rightarrow 0$. These properties at both limiting cases are in agreement with those of the exactly solvable model, \eqref{genuine monopole} and \eqref{trivial phase}, respectively, analyzed in Section 1 if one replaces $\vec{B}(t)\rightarrow \vec{p}(t)$.

\section{Anomalous commutators induced by an external magnetic monopole}

The monopole-like object induced by adiabatic Berry's phase in momentum space, which was assumed to be a genuine Dirac-type monopole,  was originally used in the analyses of the anomalous Hall effect in ferromagnetic materials \cite{Karplus, Fang} and the spin Hall effect \cite{Hirsch} in condensed matter physics. The formalism  has been later applied  to other fields also.  The effective semi-classical equations, which 
incorporate Berry's phase  near the level crossing point, are customarily adopted as (see, for example,  \cite{Niu})
\begin{eqnarray}\label{semi-classical equation}
\dot{x}_{k}=-\Omega_{kl}(\vec{p})\dot{p}_{l} +\frac{\partial \epsilon_{n}(\vec{p})}{\partial p_{k}}, \ \
\dot{p}_{k}=-eF_{kl}(\vec{x})\dot{x}_{l} + e\frac{\partial}{\partial x_{k}}\phi(\vec{x})
\end{eqnarray}
by adding the adiabatic Berry's phase $-\Omega_{kl}\dot{p}_{l}$ to the semi-classical equations of motion as an extra induced term. This addition of the adiabatic Berry's phase, which originally  appears as a phase of the wave function,  to the equations of motion is a remarkable and bold physical step, and the semi-classical motions are treated as motions in a given monopole potential located at the level crossing point \footnote{The Dirac string singularity is not observable when Berry's phase is defined inside the exponential factor \cite{Dirac, Wu-Yang}, but the removal of the Dirac string from Berry's phase requires a careful analysis when it is taken outside the exponential factor.}. Here $\epsilon_{n}(\vec{p})$ stands for the n-th energy level in the band structure. The magnetic flux $\Omega_{kl}(\vec{p})$ of Berry's phase, which is  assumed to be a Dirac monopole located at the origin of momentum space~\cite{Dirac},  and the electromagnetic tensor $F_{kl}$ are defined by  
\begin{eqnarray}
\Omega_{kl}=\frac{\partial}{\partial p_{k}}{\cal A}_{l}-\frac{\partial}{\partial p_{l}}{\cal A}_{k}, \ \  \ F_{kl}=\frac{\partial}{\partial x_{k}}A_{l}-\frac{\partial}{\partial x_{l}} A_{k},
\end{eqnarray}
respectively.  Here we defined $p_{l}=\hbar k_{l}$ to write all the equations in terms of $p_{l}$ 
compared to the notation in \cite{Niu}, to keep track of the $\hbar$ factor in a transparent way.

It is important to examine if those equations \eqref{semi-classical equation} formally written in the phase space language are consistent.  Xiao et al.~\cite{Niu}  showed that, due to the monopole curvature
term, the conserved phase-space volume is modified to
\begin{eqnarray}\label{conserved volume}
(1+e\vec{B}\cdot\vec{\Omega})d^{3}xd^{3}p/
(2\pi\hbar)^{3}
\end{eqnarray}
with $F_{kl}=\epsilon^{klm}B_{m}$ and $\Omega_{kl}=\epsilon^{klm}\Omega_{m}$.
Partly to account for this modification of the phase space volume, they showed that the equal-time commutators in the sense of Poisson brackets are modifed by $\Omega_{kl}$ and become anomalous in  a simplified treatment of the effective Lagrangian which reproduces \eqref{semi-classical equation}. The anomalous commutator was also briefly mentioned in \cite{Fang}. We emphasize that these analyses are based on the premise that adiabatic Berry's phase is represented by a genuine Dirac monopole in momentum space.

Duval et al.~\cite{Duval} further analyzed this issue  following the formulation of Faddeev and Jackiw~\cite{Faddeev-Jackiw}, essentially considering the effective action 
\begin{eqnarray}\label{action}
S=\int dt[p_{k}\dot{x}_{k} - eA_{k}(\vec{x})\dot{x}_{k} + {\cal A}_{k}(\vec{p})\dot{p}_{k}-\epsilon_{n}(\vec{p})+e\phi(\vec{x})],
\end{eqnarray}  
from which the equations in \eqref{semi-classical equation} are derived \footnote{It is interesting that one can derive the equations of motion \eqref{semi-classical equation} from an action \eqref{action} written in terms of phase-space variables. In the present paper we analyze the mathematical aspects of \eqref{semi-classical equation} and \eqref{action} by assuming that ${\cal A}_{k}(\vec{p})\dot{p}_{k}$ stands for adiabatic Berry's phase, without asking the physical relevance of \eqref{semi-classical equation} and \eqref{action}.} ; a simplified version of this action was used in \cite{Niu}. We here assume time-independent $A_{k}$ and $\phi$, for simplicity. They
showed that a modified canonical formulation in the sense of Poisson brackets is possible by inverting a symplectic matrix defined by the action in the extended phase space. They then reproduced the conserved phase space volume \eqref{conserved volume} suggested in \cite{Niu} and also the full anomalous Poisson brackets induced by the genuine monopole curvature
\begin{eqnarray}\label{Poisson bracket} 
&&\{x_{k},x_{l}\}=\frac{\epsilon^{klm}\Omega_{m}}{1+e\vec{B}\cdot\vec{\Omega}} , \ \ \ \{p_{k},x_{l}\}=-\frac{\delta_{kl}+e\Omega_{k}B_{l}}{1+e\vec{B}\cdot\vec{\Omega}},\nonumber\\
&&\{p_{k},p_{l}\}=- \frac{\epsilon^{klm}eB_{m}}{1+e\vec{B}\cdot\vec{\Omega}},
\end{eqnarray}
where the factors containing $\Omega_{k}$ are anomalous.
  
This action \eqref{action} was later analyzed using the Bjorken-Johnson-Low (BJL) prescription~\cite{BJL} and a path integral formalism~\cite{fujikawa-prd2018}. This approach, which is based on the evaluation of correlation functions first and then extract equal-time commutators from them by the BJL method, is known to work in various applications including all the known anomalous commutators associated with quantum anomalies. 
The path integral analysis used in~\cite{fujikawa-prd2018} is based on the quadratic expansion of the action \eqref{action} around a classical solution in the phase space $(\vec{x}_{(0)},\vec{p}_{(0)})$ by replacing $(\vec{x},\vec{p})\rightarrow (\vec{x}_{(0)},\vec{p}_{(0)})+(\vec{x},\vec{p})$,
\begin{eqnarray}\label{action2}
S&=&\int dt[p_{k}\dot{x}_{k} -\frac{e}{2}F_{lk}(\vec{x}_{(0)})x_{l}\dot{x}_{k} + \frac{1}{2}\Omega_{lk}(\vec{p}_{(0)})p_{l}\dot{p}_{k}\nonumber\\
&& \hspace{0.5cm} -\frac{\vec{p}^{2}}{2m}+\frac{e}{2}\partial_{k}\partial_{l}\phi(\vec{x}_{(0)})x_{k}x_{l}].
\end{eqnarray}
To be definite we choose the kinetic energy term $\epsilon_{n}=\vec{p}^{2}/2m$.
Some of the lower order commutators are given by the path integral and the BJL prescription \footnote{We use the same notation $x_{k}(t)$, for example, for both classical and quantum variables, but our notational convention will not cause any confusions in the present paper.}  (on the understanding that $kl$ matrix element on the right-hand sides is taken)~\footnote{To compare the results \eqref{commutator} with Poisson brackets in \eqref{Poisson bracket}, one may use an identity $(\frac{1}{1-e\Omega F})_{kl}=(\delta_{kl}+eB_{k}\Omega_{l})/(1+e(B_{m}\Omega_{m}))$ which is valid when one defines $F_{kl}=\epsilon^{klm}B_{m}$ and $\Omega_{kl}=\epsilon^{klm}\Omega_{m}$. Similarly, $(\frac{1}{1-eF\Omega})_{kl}=(\delta_{kl}+e\Omega_{k}B_{l})/(1+e(B_{m}\Omega_{m}))$.}
\begin{eqnarray}\label{commutator}
&&[x_{k},x_{l}]=i\hbar\frac{1}{1-e\Omega F}\Omega, \ \ \ [p_{k},x_{l}]=-i\hbar\frac{1}{1-eF\Omega},\nonumber\\
&&[p_{k},p_{l}]=-i\hbar eF\frac{1}{1-e\Omega F}.
\end{eqnarray}
These commutators are expected to be generic in the spirit of the background field method. In fact, these commutators are known to be the conventional canonical commutators in the presence of the electromagnetic potential  if one sets $\Omega=0$; the evaluation in the infinitesimal neighborhood of $(\vec{x}_{(0)},\vec{p}_{(0)})$ in phase space using the action 
\eqref{action} with ${\cal A}_{k}(\vec{p})=0$ gives a reliable estimate in accord with the idea of background field method. Also, commutators with $F=0$ in \eqref{commutator} are formally exact up to the issue of non-locality, as will be shown later. But it was recognized that there appear an infinite tower of commutators in addition to those in \eqref{commutator}, which shows that the dynamics of  \eqref{action2} with a Dirac monopole in momentum space is  non-local in time. The non-locality  is seen when one first solves the equation
\begin{eqnarray}\label{non-locality}
\dot{x}_{k}=-\Omega_{kl}\dot{p}_{l} +p_{k}/m
\end{eqnarray}
in terms of $p_{k}$ for given $\dot{x}_{k}$ to write the action \eqref{action2} in  coordinate variables  $x_{k}(t)$ only. If one attempts to invert \eqref{non-locality} at a specific point $\vec{p}_{(0)}$ with $\Omega_{kl}=\Omega_{kl}(\vec{p}_{(0)})$, one obtains 
\begin{eqnarray}
p_{l}&=&\frac{1}{-\Omega_{kl}\frac{d}{dt}+\frac{1}{m}\delta_{kl}}\dot{x}_{k}\nonumber\\
&=&m\dot{x}_{l}+m^{2}\Omega_{kl}\frac{d}{dt}\dot{x}_{k} + ....
\end{eqnarray} 
and thus one has an infinite series of higher 
time-derivative terms, which shows that the theory is nonlocal in time. The non-locality shows that the formal canonical formalism of \eqref{action} postulated as an ansatz in the presence of a Dirac monopole in momentum space is not defined in the Lagrangian formalism. This complication may be interpreted that the quantization in the presence of a Dirac magnetic monopole in momentum space has  a certain intrinsic ambiguity, although those commutators shown in \eqref{commutator} are free of these complications. To the extend that one discusses only the limited commutators in \eqref{commutator} in the presence of the Dirac monopole, one may forgo the analysis of nonlocality issue.

These commutators \eqref{commutator} containing the magnetic flux $\Omega$ of a Dirac monopole, which are anomalous in the conventional sense, are supported by the quantum interpretation of Poisson brackets \eqref{Poisson bracket} given by the modified canonical formalism~\cite{Duval}. This agreement is assuring since the Poisson bracket formalism is more flexible as it enjoys the full power of classical canonical transformations
compared to the quantum mechanical analysis.
   
In the recent interest in the topological phenomena in a wider area of physics, such as in condensed matter and nuclear physics and related fields, the results of the papers in \cite{Niu, Duval}, in particular, the appearance of a genuine Dirac monopole as idealized Berry's phase and the modified commutators in the form of Poisson brackets, have been used by many authors, although we forgo mentioning those specific applications. Our main interest is to understand better the basis of the analyses in  \cite{Niu, Duval}, in particular, the deformation of the principle of quantum mechanics by Berry's phase  in the sense of anomalous canonical commutators. 

One may summarize the analysis in this section as follows: If one assumes that adiabatic Berry's phase in momentum space, which is treated as a well-defined (externally given) Dirac monopole, is  incorporated as a term in the action \eqref{action}, then the semi-classical equations of motion \eqref{semi-classical equation}, the modified phase space volume \eqref{conserved volume} and the anomalous Poisson brackets \eqref{Poisson bracket} are naturally derived. Some of these modified Poisson brackets are replaced by the modified quantum mechanical commutators in certain limiting cases as discussed above in connection with \eqref{commutator}. The action \eqref{action} with a genuine Dirac monopole in momentum space is non-local in time when converted to a Lagrangian formalism written in terms of coordinate variables only.

\section{No anomalous canonical commutators induced by Berry's phase}

The commonly assumed anomalous Poisson brackets \eqref{Poisson bracket} or  corresponding anomalous canonical commutators \eqref{commutator} defined for a genuine Dirac monopole in momentum space, if they should be confirmed to be valid for generic Berry's phase also, would imply that Berry's phase deforms the principle of quantum mechanics. This deformation of quantum mechanics would be surprising, since Berry's phase universally appears for level crossing problems in the adiabatic approximation. 
The topological aspect of Berry's phase is understood by examining the exact solution of a soluble model in Sections 1, which is confirmed by a generic model in Section 2; namely, the topology of the Dirac monopole appears  in the precise adiabatic limit while the monopole-like topology disappears in the non-adiabatic limit. On the other hand, the principle of quantum mechanics in the sense of  canonical commutators is expected to be valid regardless of adiabatic or non-adiabatic motions.
In the present section, we are going to argue that the action \eqref{action} describes a genuine Dirac monopole in momentum space in a neat manner but does not describe the full picture of Berry's phase.

To analyze this issue, we start with some preparations.
In our analysis of anomalous canonical commutators, the commutator $[x_{k},x_{l}]$ plays a central role since it is precisely defined as the quantum mechanical commutator for a genuine Dirac monopole if electromagnetic potentials are absent in \eqref{action}. We can thus make a more definite statement on the consequence of \eqref{action} than the classical Poisson bracket.  We thus examine the action without electromagnetic potentials obtained from \eqref{action} by setting $A_{k}=\phi=0$,
\begin{eqnarray}\label{action3}
S=\int dt[p_{k}\dot{x}_{k} + {\cal A}_{k}(\vec{p})\dot{p}_{k}-\epsilon_{n}(\vec{p})].
\end{eqnarray}  
This action in the case of an externally given monopole potential is quantized exactly by rewriting it as
\begin{eqnarray}\label{action04}
S=\int dt[-\dot{p}_{k}X_{k}- \epsilon_{n}(\vec{p})]=\int dt[p_{k}\dot{X}_{k}- \epsilon_{n}(\vec{p})]
\end{eqnarray}
defining an auxiliary variable (this change of variables is natural in the path integral)
\begin{eqnarray}\label{change of variables}
X_{k}(t)=x_{k}(t)- {\cal A}_{k}(\vec{p}).
\end{eqnarray} 
The action \eqref{action04}, when regarded as a functional of two variables $X_{k}$ and $p_{k}$ with $p_{k}=\frac{\delta S}{\delta \dot{X}_{k}}$,  gives rise to the canonical commutators
\begin{eqnarray}\label{free commutator1}
[X_{k},X_{l}]=0, \ \ \ [X_{k},p_{l}]=i\hbar\delta_{kl}, \ \ [p_{k},p_{l}]=0,
\end{eqnarray}
which are equivalent to 
\begin{eqnarray}\label{exact canonical commutator}
[x_{k},x_{l}]=i\hbar \Omega_{kl}, \ \ \ [x_{k},p_{l}]=i\hbar\delta_{kl}, \ \ [p_{k},p_{l}]=0
\end{eqnarray}
when one uses \eqref{change of variables}. This exact quantization in the phase space approach agrees with the replacement of the Poisson brackets given in \eqref{Poisson bracket} by canonical commutators and also with our results in \eqref{commutator} if one sets electromagnetic fields to be zero. We do not recognize the non-locality issue in the phase space approach to the canonical quantization. 

We next confirm that Berry's phase in momentum space \eqref{adiabatic Stone phase3} is an  induced $O(\hbar)$ quantum mechanical effect with a quantized magnetic charge $e_{M}=2\pi\hbar$ in \eqref{monopole charge} since it arises from the analysis of the Schr\"{o}dinger equation  \eqref{Schroedinger equation}. 
This value of the magnetic charge is related to the Dirac quantization condition \cite{Dirac} and more directly to the Wu-Yang gauge invariance condition of the physical amplitude in the presence of a magnetic monopole-like potential \cite{Wu-Yang}.
In the analysis of Wu and Yang, the Aharonov-Bohm phase was used as a basic 
physical quantity. In the present case, we have the Schr\"{o}dinger amplitude \eqref{eq-exactamplitude1} as the physical amplitude and its invariance under the gauge transformation is fundamental \cite{Deguchi-Fujikawa-2019}.

We briefly explain the reason for $e_{M}=2\pi\hbar$ in connection with the Dirac quantization condition. 
We analyze the adiabatic configuration with $\eta=\mu TB/\pi >1$ namely $B>\pi/\mu T$ such as in Fig.2.  The argument of Wu and Yang is to consider the singularity-free potentials in the upper and lower hemispheres  with radius $B$      
\begin{eqnarray}
{\cal A}_{\varphi +} 
&=&  \frac{e_{M}}{4\pi B\sin\theta}(1 - \cos\Theta(\theta)),\nonumber\\
{\cal A}_{\varphi -} 
&=&  \frac{e_{M}}{4\pi B\sin\theta}(- 1 - \cos\Theta(\theta)),
\end{eqnarray}
using the potential in \eqref{potential1}. 
These two potentials are related by a gauge transformation 
\begin{eqnarray}
{\cal A}_{\varphi -}={\cal A}_{\varphi +} - \frac{\partial\Lambda}{B\sin\theta\partial \varphi}
\end{eqnarray}
with 
\begin{eqnarray}
\Lambda=\frac{e_{M}}{2\pi} \varphi.
\end{eqnarray}
The physical condition near the equator is
\begin{eqnarray}
\exp[-\frac{i}{\hbar}\oint {\cal A}_{\varphi -}B\sin\theta d\varphi]&=&\exp[-\frac{i}{\hbar}\oint{\cal A}_{\varphi +}B\sin\theta d\varphi +\frac{i}{\hbar}\oint \frac{\partial\Lambda}{B\sin\theta\partial \varphi}B\sin\theta d\varphi]\nonumber\\
&=&\exp[-\frac{i}{\hbar}\oint{\cal A}_{\varphi +}B\sin\theta d\varphi]
\end{eqnarray}
which is in fact satisfied since  for $e_{M}=2\pi \hbar$ the gauge term gives
\begin{eqnarray}
\exp[i e_{M}/\hbar]=\exp[2\pi i]=1
\end{eqnarray}
and thus defines a monopole-like configuration for $\eta=\mu TB/\pi >1$, i.e.,     $B>\pi/\mu T$. 
It is known that the present argument of gauge transformation is equivalent to the evaluation of the phase change induced by the Dirac string in the case of an electron placed in the field of the genuine Dirac monopole~\cite{Wu-Yang}. 
The fact that the physical condition is satisfied 
shows that the magnetic charge of Berry's phase
\begin{eqnarray}\label{magnetic charge}
e_{M}=2\pi \hbar
\end{eqnarray}
is properly quantized satisfying the Dirac quantization condition, although we have no analogue of an electric coupling in the present case unlike the original Dirac monopole~\cite{Dirac}. The specific value $e_{M}=2\pi \hbar$ is naturally valid for the magnetic monopole given by Berry's phase in the precise adiabatic limit $T\rightarrow \infty$.

Using those preparatory analyses, we first argue that the modification of the phase space volume \eqref{conserved volume} {\em in the adiabatic limit} and the anomalous canonical commutators are logically independent in principle. We then argue that the deformation of quantum mechanics in the form of the anomalous canonical commutators \eqref{exact canonical commutator} is not induced by generic Berry's phase in momentum space:

(i) We begin with the semi-classical equations of motion \eqref{semi-classical equation} with induced Berry's phase, which is the starting point of our analysis. We have
\begin{eqnarray}\label{equation2}
\dot{x}_{k}&=&-\Omega_{kl}(p(t); \mu|p|T/\pi)\dot{p}_{l}(t) +\frac{\partial \epsilon_{n}(p)}{\partial p_{k}}, \ \
\dot{p}_{k}=-eF_{kl}\dot{x}_{l} +e\frac{\partial}{\partial x_{k}}\phi(\vec{x})
\end{eqnarray}
if one uses the precise specification of the monopole-like potential \eqref{Berry connection}, which is also supported by an exactly solvable model, instead of the first equation in \eqref{semi-classical equation} with a genuine Dirac monopole; $T$ stands for the typical time scale of the system.
If one chooses a suitable $T$ which represents the time scale of the system accurately, the equations of motion are accurate. In the adiabatic limit $\mu|\vec{p}|T\gg 2\pi$ which implies a very slow movement (or oscillations) of the electron,  it is accurate to use the Dirac monopole $\Omega_{kl}(p(t))\dot{p}_{l}(t)$ in place of Berry's phase.
In the non-adiabatic limit $\mu|\vec{p}|T\ll 2\pi$, on the other hand, Berry's phase disappears and thus the equations of motion without Berry's phase are accurate.   In the equations of motion,  we assume that   {\em one will in principle be able to  control the movement of particles experimentally}. The successful applications of Berry's phase on the basis of \eqref{semi-classical equation} with emphasis on adiabatic approximation \cite{Karplus, Fang, Hirsch} are thus consistent with our analysis of Berry's phase in Sections 1 and 2. 

The effective semi-classical equations of motion \eqref{semi-classical equation}  are the classical equations with the monopole-like potential defined  in the {\em adiabatic limit} added, as explained above. To maintain the consistency, one thus needs to modify the phase space volume accordingly.
 The required modified phase space volume in \eqref{conserved volume} 
 \begin{eqnarray}\label{modified volume2}
 (1+e\vec{B}\cdot\vec{\Omega} )\frac{d^{3}xd^{3}p}{(2\pi\hbar)^{3}}, 
\end{eqnarray}
 which was introduced by Xiao et al.~\cite{Niu},  is understood as a result of the inclusion of the adiabatic Berry's phase, that is assumed to be of a genuine monopole form, into the classical equations of motion in the  adiabatic limit. 
The modification of the phase space volume \eqref{modified volume2} by Berry's phase in the context of equations of motion is thus valid in the {\em very restricted adiabatic motion in phase space}  where the adiabatic approximation to define a genuine monopole-like Berry's phase is  valid; as already noted, we are assuming that the movement of the electron (very fast or very slow, for example) in the equations of motion is in principle controlled in experiments. 
 
 We thus understand the reason why the modification of the phase space volume  is physically relevant, {\em without referring to} the modification of canonical commutators of which analysis will be shown later to be beyond the scope of adiabatic approximation.\footnote{If one should assume the genuine magnetic monopole form of Berry's phase and its validity regardless in either adiabatic or nonadiabatic domain, that is however not supported by the precise treatment of Berry's phase in Sections 1 and 2, both the modified phase space volume and anomalous canonical commutators would be derived using the action \eqref{action} as is shown in \cite{Niu,Duval}.} 

(ii) We next discuss the canonical commutators.  
In the assumed canonical commutators with a genuine monopole in \eqref{commutator}, which are supported by the general Poisson brackets \eqref{Poisson bracket}  in a semi-classical context, the terms with $1$ and $F$ are order $O(\hbar)$ as they should be (and in fact we know that they are exact if $\Omega=0$), but the terms with $\Omega$ such as (setting $F=0$ to make the analysis precise)
\begin{eqnarray}\label{anomalous coordinate commutator}
[x_{k},x_{l}]=i\hbar\Omega_{kl}
\end{eqnarray}
are of order $O(\hbar^{2})$, since $\Omega$ is of order $O(\hbar)$. In comparison, the {\em externally given} genuine Dirac monopole would give rise to the right-hand side of \eqref{anomalous coordinate commutator} with order $O(\hbar)$ and thus would be consistent with the canonical quantization in the presence of a Dirac monopole in momentum space. Apparently, \eqref{anomalous coordinate commutator} is not understood as a part of canonical commutators for the physical system which gives rise to Berry's phase. 

This $O(\hbar^{2})$ effect appears firstly because Berry's phase is assumed to be of the genuine Dirac monopole form. Secondly, the quantum mechanically induced adiabatic Berry's phase of an assumed  genuine Dirac monopole form is included in the supposed to be the classical action \eqref{action} to define quantization.  Berry's phase takes the genuine monopole form only in the precise adiabatic motion of $\vec{p}(t)$, but Berry's phase of the genuine Dirac monopole, once included in the effective action, is assumed to be valid for any time-dependent motion of $\vec{p}(t)$, either adiabatic or nonadiabatic. One cannot control the movement of dynamical variables inside the action, unlike the movement of dynamical variables in equations of motion \eqref{equation2}.\footnote{A related important fact is that an effective action, from which the semi-classical equations of motion in \eqref{equation2} with generic Berry's phase would be derived,  {\em is not defined in general}  since it is not easy to incorporate the notion of typical time scale $T$ into the action written in the t-representation.} The appearance of the non-canonical commutators with $\Omega$ is a result of the assumption that Berry's phase has a genuine Dirac monopole form for any movement.

To avoid the canonical commutators with $O(\hbar^{2})$, we propose to define quantization using the original classical Lagrangian \eqref{action} {\em without} the Dirac monopole term which can represent the quantum mechanically induced Berry's phase only in the precise adiabatic limit.  After all, Berry's phase is not an externally given monopole but rather a self-generated quantum mechanical effect in the quantized system, as is seen in the analyses in Sections 1 and 2.  An additional merit of this proposed quantization procedure is that we can avoid the action non-local in time \eqref{non-locality}.
  
The above quantization procedure we propose is consistent with  how Berry's phase is derived  in the Born-Oppenheimer approximation of the Schr\"{o}dinger equation~\cite{Higgins, Baer}; one performs quantization of both fast and slow variables  
in the master Schr\"{o}dinger equation only once. One thus obtains only   the  normal canonical commutators for slow variables {\em without} the additional induced quantum mechanical quantity such as  $\Omega$.  The above proposed quantization procedure also appears to be natural in view of the general formulation of quantum field theory. For example, the induced anomalous 
magnetic moment, which may be regarded as analogous to Berry's phase in the present problem, is used in the phenomenology of QED such as the description of a semi-classical motion of spin inside the external magnetic field but the quantization of QED is performed only once by the original Lagrangian. 

We thus conclude that the normal Poisson brackets \eqref{Poisson bracket} or canonical commutators \eqref{commutator} {\em without} the assumed contribution $\Omega$ of the Dirac monopole are the natural and correct ones in the presence of generic Berry's phase. The monopole-like Berry's phase which appears in the adiabatic approximation of the Schr\"{o}dinger equation does not deform the principle of quantum mechanics. 

\section{Order $O(\hbar^{2})$ corrections to commutators}

One may wonder if one can understand the extra terms of order $O(\hbar^{2})$ in commutators \eqref{anomalous coordinate commutator} or Poisson brackets  as  higher order corrections in quantum mechanics and thus consistent with the principle of quantum mechanics. 
We shall show that this view could be valid if one should assume that the genuine Dirac monopole form of adiabatic Berry's phase is valid even in the non-adiabatic limit,  but in reality such a view is not possible if one uses the more accurate form of Berry's phase as in \eqref{Berry connection}. This analysis will illustrate that  the possible corrections to  canonical commutators are determined by the behavior of Berry's phase at the non-adiabatic limit.

To discuss this issue, the Bjorken-Johnson-Low (BJL) prescription~\cite{BJL}, which analyzes the equal-time commutators from a point of view of the operator product at the short-time limit, is convenient \footnote{In the canonical quantization, it is not easy to see clearly that one is actually using the action \eqref{action} for rapidly oscillating non-adiabatic motions to decide equal-time commutators.}.  
The BJL prescription is known to work for all the known anomalous commutators associated with quantum anomalies. 
 In this prescription, we analyze the short-time behavior of the time ordered operator product~\footnote{We indicate the time-ordering operation by $\hat{T}$ to avoid the confusion with the period $T$. }
$\langle \hat{T} x_{k}(t)x_{l}(0)\rangle$
or large frequency limit when Fourier transformed to define the equal-time commutator in the manner~\cite{BJL, fujikawa-prd2018}
\begin{eqnarray}\label{BJL2}
&&\lim_{\omega\rightarrow\infty}-i\omega \int_{-\infty}^{\infty} dt e^{i\omega t}\langle \hat{T} x_{k}(t)x_{l}(0)\rangle\nonumber\\
&&=\lim_{\omega\rightarrow\infty}\int_{-\infty}^{\infty} dt e^{i\omega t}\frac{d}{dt}\langle \hat{T}x_{k}(t)x_{l}(0)\rangle \nonumber\\
&&=\lim_{\omega\rightarrow\infty}\int_{-\infty}^{\infty} dt e^{i\omega t}\{\langle \hat{T} \dot{x}_{k}(t)x_{l}(0)\rangle+\delta(t)[x_{k}(0),x_{l}(0)]\}\nonumber\\
&&=[x_{k}(0),x_{l}(0)]
\end{eqnarray}
where the first term in the third line, which is well-defined at $t=0$ by the definition of $\hat{T}$ product, vanishes 
for $\omega\rightarrow \infty$ due to Riemann-Lebesgue lemma. 

Since the BJL prescription is not commonly used in quantum mechanics, we illustrate the definition of the time-ordered product by the phase-space path integral
\begin{eqnarray}
\langle \hat{T} x_{k}(t_{1})x_{l}(t_{2})\rangle=\int {\cal D}\vec{x}{\cal D}\vec{p} x_{k}(t_{1})x_{l}(t_{2})\exp \{\frac{i}{\hbar}S \}
\end{eqnarray}
for a free particle
\begin{eqnarray}\label{free particles}
S=\int dt [p_{k}(t)\dot{x}_{k}(t) -\frac{1}{2m}p^{2}_{k}(t)].
\end{eqnarray}
Following the prescription of path integral, we obtain
\begin{eqnarray}\label{Green's function}
\langle \hat{T} x_{k}(t_{1})x_{l}(t_{2}) \rangle&=& \frac{i\hbar}{m}
\int \frac{d\omega}{2\pi}\frac{\delta_{kl}}{\omega^{2}}e^{i\omega(t_{1}-t_{2})}\nonumber\\
&=&\frac{i\hbar}{m}
\int \frac{d\omega}{2\pi}\frac{\delta_{kl}}{\omega^{2}-\tilde{\mu}^{2}}e^{i\omega(t_{1}-t_{2})},\nonumber\\
\langle \hat{T} x_{k}(t_{1})p_{l}(t_{2}) \rangle &=&
\hbar\int \frac{d\omega}{2\pi}\frac{\delta_{kl}}{\omega}e^{i\omega(t_{1}-t_{2})}\nonumber\\
&=&\hbar
\int \frac{d\omega}{2\pi}\frac{\omega\delta_{kl}}{\omega^{2}-\tilde{\mu}^{2}}e^{i\omega(t_{1}-t_{2})}
\end{eqnarray}
and $\langle \hat{T} p_{k}(t_{1})p_{l}(t_{2}) \rangle = 0$. 
The standard procedure to derive these correlation functions is to add the source terms to the action \eqref{free particles}
\begin{eqnarray}
S=\int dt [p_{k}(t)\dot{x}_{k}(t) -\frac{1}{2m}p^{2}_{k}(t)+ x_{k}(t)J^{x}_{k}(t)+p_{k}(t)J^{p}_{k}(t)]
\end{eqnarray}
and perform the quadratic path integral. Taking the functional derivatives with respect to source terms, one then obtains the correlations functions. One can also understand the explicit expression of the first relation $\langle \hat{T} x_{k}(t_{1})x_{l}(t_{2}) \rangle$ in \eqref{Green's function} using the free action $S=\int dt [\frac{m}{2}\dot{x}_{k}(t)^{2}]$, which is analogous to a free Klein-Gordon particle with vanishing mass and momentum, by first performing the path integral over momentum variables $p_{k}$.
In \eqref{Green's function} we introduced a generalization of Feynman's  $i\epsilon$-prescription by 
\begin{eqnarray}
\tilde{\mu}^{2} \equiv \mu^{2}-i\epsilon
\end{eqnarray}
using two infinitesimal positive constants $\mu$
and $\epsilon$ by choosing $\mu\gg \epsilon$ for a technical reason. This Feynman's prescription specifies the time-ordered product, for example,
\begin{eqnarray}
\langle \hat{T} x_{k}(t_{1})x_{l}(t_{2}) \rangle = \langle x_{k}(t_{1})x_{l}(t_{2})  \rangle\theta(t_{1} - t_{2}) + \langle x_{l}(t_{2})x_{k}(t_{1}) \rangle \theta(t_{2} - t_{1}).
\end{eqnarray} 
We then obtain from \eqref{Green's function}
\begin{eqnarray}
&&(-i\omega)\int dt_{1} e^{i\omega(t_{1}-t_{2})}\langle \hat{T} x_{k}(t_{1})x_{l}(t_{2}) \rangle \nonumber\\
&=& \int dt_{1} e^{i\omega(t_{1}-t_{2})}\frac{d}{dt}\langle \hat{T} x_{k}(t_{1})x_{l}(t_{2}) \rangle\nonumber\\
&=& \int dt_{1} e^{i\omega(t_{1}-t_{2})}\{\langle \hat{T} \dot{x}_{k}(t_{1})x_{l}(t_{2}) \rangle +[x_{k}(t_{1}), x_{l}(t_{2})]\delta(t_{1}-t_{2})\}\nonumber\\
&=& (-i\omega)\frac{i\hbar}{m}
\frac{1}{\omega^{2}-\tilde{\mu}^{2}}.
\end{eqnarray}
By analyzing the large $\omega$ behavior, we conclude 
\begin{eqnarray}\label{commutator1}
[x_{k}(t_{1}), x_{l}(t_{1})] = 0
\end{eqnarray}
and  
\begin{eqnarray}
\int dt_{1} e^{i\omega(t_{1}-t_{2})}\{\langle \hat{T} \dot{x}_{k}(t_{1})x_{l}(t_{2}) \rangle =  (-i\omega)\frac{i\hbar}{m}
\frac{\delta_{kl}}{\omega^{2}-\tilde{\mu}^{2}},
\end{eqnarray}
which in turn implies by repeating the above procedure 
\begin{eqnarray}
&&(-i\omega)\int dt_{1} e^{i\omega(t_{1}-t_{2})}\{\langle \hat{T} \dot{x}_{k}(t_{1})x_{l}(t_{2}) \rangle \nonumber\\
&=&
\int dt_{1} e^{i\omega(t_{1}-t_{2})}\{\langle \hat{T} \ddot{x}_{k}(t_{1})x_{l}(t_{2}) \rangle +[\dot{x}_{k}(t_{1}), x_{l}(t_{2})]\delta(t_{1}-t_{2})\}\nonumber\\
&=&  (-i\omega)^{2}\frac{i\hbar}{m}
\frac{\delta_{kl}}{\omega^{2}-\tilde{\mu}^{2}}.
\end{eqnarray}
By analyzing the limit $\omega\rightarrow \infty$, we conclude
\begin{eqnarray}
[\dot{x}_{k}(t_{1}), x_{l}(t_{1})] = \frac{-i\hbar}{m}\delta_{kl}
\end{eqnarray}
which gives the conventional relation $[m\dot{x}_{k}(t_{1}), x_{l}(t_{1})] = [p_{k}(t_{1}), x_{l}(t_{1})] =-i\hbar\delta_{kl}$, and $\langle \hat{T} \ddot{x}_{k}(t_{1})x_{l}(t_{2}) \rangle=0$ uisng the equations of motion.

Similarly, one obtains from the second relation in \eqref{Green's function}
\begin{eqnarray}
&&(-i\omega)\int dt_{1} e^{i\omega(t_{1}-t_{2})}\langle \hat{T} x_{k}(t_{1})p_{l}(t_{2}) \rangle \nonumber\\
&=& \int dt_{1} e^{i\omega(t_{1}-t_{2})}\frac{d}{dt}\langle \hat{T} x_{k}(t_{1})p_{l}(t_{2}) \rangle\nonumber\\
&=& \int dt_{1} e^{i\omega(t_{1}-t_{2})}\{\langle \hat{T} \dot{x}_{k}(t_{1})p_{l}(t_{2}) \rangle +[x_{k}(t_{1}), p_{l}(t_{2})]\delta(t_{1}-t_{2})\}\nonumber\\
&=& (-i\omega)
\frac{-\hbar\omega\delta_{kl}}{\omega^{2}-\tilde{\mu}^{2}}
\end{eqnarray}
which gives at the large $\omega$ limit
\begin{eqnarray}\label{commutator2}
[x_{k}(t_{1}), p_{l}(t_{2})] =i\hbar\delta_{kl}
\end{eqnarray}
and 
$\langle \hat{T} \dot{x}_{k}(t_{1})p_{l}(t_{2}) \rangle= 0 =\langle \hat{T} p_{k}(t_{1})p_{l}(t_{2}) \rangle$ in the sense of $\hat{T}$-product.
We also obtain
\begin{eqnarray}\label{commutator3}
[p_{k}(t_{1}), p_{l}(t_{1})]=0
\end{eqnarray}
from $\langle \hat{T} p_{k}(t_{1})p_{l}(t_{2}) \rangle = 0$.  We thus obtain the standard results of the canonical quantization in \eqref{commutator1}, \eqref{commutator2} and  \eqref{commutator3}
\footnote{ These examples show that we can formally ignore Feynman's $i\epsilon$ prescription in the analysis of BJL limit. We use this simplified treatment of the BJL limit in the present paper.}.

\subsection{Dirac monopole case}
 We obtain the relations \eqref{free commutator1} by the present path integral \eqref{free particles} if one replaces $x_{k}(t)\rightarrow X_{k}(t)$.  
Now the relations \eqref{exact canonical commutator} are evaluated by the path integral, for example,
\begin{eqnarray}
\langle\hat{T} x_{k}(t_{1})x_{l}(t_{2})\rangle =
\int {\cal D}X{\cal D}p x_{k}(t_{1})x_{l}(t_{2}) \exp\{\frac{i}{\hbar}S_{J}\}
\end{eqnarray}
with 
\begin{eqnarray}\label{free Lagrangian2}
S_{J}=\int dt [p_{k}(t)\dot{X}_{k}(t) -\frac{1}{2m}p^{2}_{k}(t)+ X_{k}(t)J_{k}(t)]
\end{eqnarray}
 and $x_{k}(t)=X_{k}(t) + {\cal A}(p_{k}(t)$. 
Namely,
\begin{eqnarray}
\langle\hat{T} x_{k}(t_{1})x_{l}(t_{2})\rangle_{J}&=&\langle\hat{T} X_{k}(t_{1})X_{l}(t_{2})\rangle_{J}+\langle\hat{T} X_{k}(t_{1}){\cal A}(p_{l}(t_{2}))\rangle_{J} \nonumber\\
&&+
\langle\hat{T} {\cal A}(p_{l}(t_{1}))X_{l}(t_{2})\rangle_{J}
\end{eqnarray}
by noting $\langle \hat{T} p_{k}(t_{1})p_{l}(t_{2}) \rangle = 0$.  The path integral analysis, which corresponds to Wick's prescription, gives 
\begin{eqnarray}
\langle\hat{T} X_{k}(t_{1}){\cal A}(p_{l}(t_{2}))\rangle_{J} = \langle\hat{T} X_{k}(t_{1})p_{l}(t_{2})\rangle_{J}\langle \frac{\delta}{\delta p_{l}(t_{2})}{\cal A}(p_{l}(t_{2}))\rangle_{J}.
\end{eqnarray}
 If one retains only the connected components,
\begin{eqnarray}
\langle\hat{T} X_{k}(t_{1})X_{l}(t_{2})\rangle_{J}&=&\langle\hat{T} X_{k}(t_{1})X_{l}(t_{2})\rangle,\nonumber\\
\langle\hat{T} X_{k}(t_{1})p_{l}(t_{2})\rangle_{J}\langle \frac{\delta}{\delta p_{l}(t_{2})}{\cal A}(p_{l}(t_{2}))\rangle_{J}&=&\langle\hat{T} X_{k}(t_{1})p_{l}(t_{2})\rangle\langle \frac{\delta}{\delta p_{l}(t_{2})}{\cal A}(p_{l}(t_{2}))\rangle_{J}.
\end{eqnarray}
By applying the BJL analysis to $\langle\hat{T} X_{k}(t_{1})X_{l}(t_{2})\rangle$ and 
$\langle\hat{T} X_{k}(t_{1})p_{l}(t_{2})\rangle$ for the free Lagrangian \eqref{free Lagrangian2}, one thus obtains
\begin{eqnarray}\label{anomalous commutator1}
\langle [x_{k}(t_{1}),x_{l}(t_{1})]\rangle_{J} &=&[X_{k}(t_{1}),X_{l}(t_{1})]\nonumber\\
&&+ i\hbar \langle \frac{\delta}{\delta p_{k}(t_{1})}{\cal A}_{l}(p_{l}(t_{1}))- \frac{\delta}{\delta p_{l}(t_{1})}{\cal A}_{k}(p_{l}(t_{1}))\rangle_{J}\nonumber\\
&=& i\hbar \langle \Omega_{kl}(p(t_{1}))\rangle_{J}
\end{eqnarray}
namely, one reproduces the result of the operator analysis
\begin{eqnarray}
[x_{k}(t_{1}),x_{l}(t_{1})] = i\hbar \Omega_{kl}(p(t_{1})).
\end{eqnarray}
Other terms in \eqref{exact canonical commutator} in the case of the genuine Dirac monopole are similarly derived. These analyses illustrate how to derive the operator formalism from the path integral fromalism.

\subsection{Generic Berry's phase}
We analyze the case with generic Berry's phase by considering the infinitesimal neighborhood of the phase space point $(\vec{x}_{(0)},\vec{p}_{(0)})$ in the spirit of the background field method.  The classical solutions of \eqref{action} for  $A_{k}(\vec{x})=\phi(\vec{x})=0$ are
\begin{eqnarray}\label{classical solution}
\frac{d}{dt} \vec{p}_{(0)} &=& 0,\nonumber\\
\frac{d}{dt} \vec{x}_{(0)} &=& \frac{\partial \epsilon_{n}(p)}{\partial \vec{p}_{(0)}} 
\end{eqnarray}
which are independent of the form of Berry's phase.
We thus choose those classical solutions to define the background field method and replace $\vec{x}\rightarrow \vec{x}_{(0)} + \vec{x}$ and $\vec{p}\rightarrow \vec{p}_{(0)} + \vec{p}$ with time independent $\vec{p}_{(0)}$.
We then approximate 
Berry's phase 
${\cal A}_{k}(\vec{p}; 2\hbar\mu|\vec{p}| /\hbar\omega )\dot{p}_{k}$ in  \eqref{Berry connection} by
\begin{eqnarray}
\int dt {\cal A}_{k}(\vec{p}; 2\hbar\mu|\vec{p}| /\hbar\omega)\dot{p}_{k}&=&\int dt{\cal A}_{k}(\vec{p}_{(0)}+\vec{p}; 2\hbar\mu|\vec{p}_{(0)}| /\hbar\omega)\dot{p}_{k}\nonumber\\
&\simeq& \int dt \frac{1}{2}p_{l}(t) \Omega_{lk}(\vec{p}_{(0)}, \partial_{t})\dot{p}_{k}(t) 
\end{eqnarray}
where we set the typical magnitude of momentum
in $\eta=2\hbar\mu|\vec{p}| /\hbar\omega$  at $\vec{p}=\vec{p}_{(0)}$ and identified $i\partial_{t} \sim \omega$. 
In this form of Berry's phase, we can incorporate it in the action 
\begin{eqnarray}\label{action4}
S=\int dt[p_{k}\dot{x}_{k} + \frac{1}{2}p_{l}(t) \Omega_{lk}(\vec{p}_{(0)}, \partial_{t})\dot{p}_{k}(t)-\frac{1}{2m}p^{2}_{k}(t)]
\end{eqnarray} 
that stands for a general class of higher derivative theory \cite{Nambu} which is treated by the formal path integral, for example, by recovering the canonical structure by the BJL method later.
The classical solutions of \eqref{action4} have the same form as the above solutions \eqref{classical solution}, and thus consistent.

We effectively introduce a ``form factor'' for the induced Berry's phase, for example, 
\begin{eqnarray}
\Omega_{lk}(\vec{p}_{(0)}, \partial_{t})\dot{p}_{k}(t)&=& \int dt^{\prime}\Omega_{lk}(\vec{p}_{(0)})\frac{\omega_{0}^{2}}{[\omega_{0}^{2}-\partial_{t}^{2}]}\delta(t-t^{\prime})\dot{p}_{k}(t^{\prime})\nonumber\\
&=&\int dt^{\prime}\int\frac{d\omega}{2\pi}\Omega_{lk}(\vec{p}_{(0)})\frac{\omega_{0}^{2}}{[\omega_{0}^{2}+\omega^{2}]}e^{i\omega(t-t^{\prime})}\dot{p}_{k}(t^{\prime})\nonumber\\
&=&\Omega_{lk}(\vec{p}_{(0)})\int\frac{d\omega}{2\pi}e^{i\omega t}\frac{\omega_{0}^{2}}{[\omega_{0}^{2}+\omega^{2}]}\int dt^{\prime}e^{-i\omega t^{\prime}}\dot{p}_{k}(t^{\prime})
\end{eqnarray}
where a simple form factor is used for illustration purposes
\begin{eqnarray}\label{form factor}
\Omega_{lk}(\vec{p}_{(0)}, \omega)=\Omega_{lk}(\vec{p}_{(0)})\frac{\omega_{0}^{2}}{[\omega_{0}^{2}+\omega^{2}]}
\end{eqnarray}
which describes the Dirac monopole $\Omega_{lk}(\vec{p}_{(0)})$ at the adiabatic limit $\omega\ll \omega_{0}$ while  
the non-adiabatic behavior \eqref{non-adiabatic form}, $\sim 1/\omega^{2}$, is realized for $\omega\gg \omega_{0}$.  
 
The auxiliary variables in \eqref{change of variables} are now defined by 
\begin{eqnarray}
X_{k}(t) = x_{k}(t) + 
\frac{1}{2}\Omega_{kl}(\vec{p}_{(0)}, \partial_{t})p_{l}(t),
\end{eqnarray}
and we evaluate $\langle\hat{T}x_{k}(t_{1})x_{l}(t_{2})\rangle$ for the action $S=\int dt [p_{k}(t)\dot{X}_{k}(t) -\frac{1}{2m}p^{2}_{k}(t)]$. We then encounter the combination
\begin{eqnarray}
\langle \hat{T} X_{k}(t_{1})\frac{1}{2}\Omega_{lm}(\vec{p}_{(0)}, \partial_{t_{2}})p_{m}(t_{2})\rangle
= \frac{\hbar}{2}\int\frac{d\omega}{2\pi}e^{i\omega(t_{1}-t_{2})}
\frac{\Omega_{lk}(p_{(0)},\omega)}{-\omega}
\end{eqnarray}
where we incorporated the factor $\frac{1}{2}\Omega_{lm}(\vec{p}_{(0)}, \partial_{t_{2}})$ into the free correlation  $\langle\hat{T}X_{k}(t_{1})p_{l}(t_{2})\rangle$. 
The BJL prescription then gives 
\begin{eqnarray}
[X_{k}(t_{1}), \frac{1}{2}\Omega_{lm}(\vec{p}_{(0)}, \partial_{t_{1}})p_{m}(t_{1})] = \lim_{\omega\rightarrow\infty} (-i\omega)\frac{\hbar}{2}\frac{\Omega_{lk}(p_{(0)},\omega)}{-\omega}.
\end{eqnarray}
Adding the contribution of another combination, one finally obtains 
\begin{eqnarray}
[x_{k}(t_{1}),x_{l}(t_{1})]=\lim_{\omega\rightarrow\infty} (-i\omega)\hbar\frac{\Omega_{kl}(p_{(0)},\omega)}{-\omega}
\end{eqnarray}
using $[X_{k}(t_{1}), X_{l}(t_{1})]=[p_{k}(t_{1}), p_{l}(t_{1})]=0$.
If one uses the specific form factor \eqref{form factor}, one concludes
\begin{eqnarray}\label{normal commutator}
[x_{k}(t_{1}),x_{l}(t_{1})]=0
\end{eqnarray}
for any finite $\omega_{0}$. 
 {\em To obtain the conventional anomalous result \eqref{anomalous coordinate commutator}, generic Berry's phase needs to be  the Dirac monopole form for all values of frequencies $\lim_{\omega\rightarrow\infty}\Omega(p_{(0)},\omega)=\Omega(p_{(0)})$.}  
 
Any form factor $\Omega_{lk}(\vec{p}_{(0)}, \omega)$ which satisfies  $\Omega_{lk}(\vec{p}_{(0)}, 0)=\Omega_{lk}(\vec{p}_{(0)})$ and vanishes for $\omega\rightarrow \infty$ leads to the same conclusion.
For a generic case \eqref{Berry connection}, one might expect a milder 
behavior
\begin{eqnarray}
\Omega_{kl}(p_{(0)},\omega) \sim \frac{1}{\omega}
\end{eqnarray}
but still one obtains the vanishing anomalous term in \eqref{normal commutator}.
We thus conclude that Berry's phase induced by quantum mechanical effects does not modify the equal-time commutators, while the assumed genuine Dirac monopole would lead to anomalous canonical commutators.

\section{Conclusion}
The precisely specified quantum mechanically induced 
Berry's phase \eqref{Berry connection} does not lead to the deformation of the principle of quantum mechanics in the sense of anomalous canonical commutators. The action \eqref{action} provides an interesting model of a genuine Dirac monopole in momentum space, but it does not describe the full picture of Berry's phase.
The important practical implication of the present analysis is that one should take the applications of adiabatic  Berry's phase, which emphasize anomalous Poisson brackets \eqref{Poisson bracket} or  anomalous canonical commutators containing $\Omega_{kl}$ in \eqref{commutator}, with due care since one is then assuming the deformation of the principle of quantum mechanics by Berry's phase.

In comparison, it is well-known that the quantum anomalies, which are  short distance phenomena, generally lead to the deformation of quantum mechanical current commutators when analyzed using the BJL prescription~\cite{Jackiw, Adler, FS}. But the deformation in this case is a quantum modification of symmetry principles from the naive expectations based on Noether's theorem and not the deformation of the principle of quantum mechanics itself \footnote{It is nevertheless  interesting that the large $\omega$ behavior of the genuine Dirac monopole in momentum space determines the possible deformation of commutator \eqref{anomalous commutator1}(and also \eqref{BJL3} later), which is analogous to the quantum anomaly appearing in the triangle diagram in the sense that the large frequency limit generates anomalous results in both cases.}.
\\

One of us (KF) is supported in part by JSPS KAKENHI (Grant No.18K03633).

\appendix
\section{Perturbative treatment of Berry's phase}

We  illustrate the treatment of Berry's phase as a perturbation on the time-ordered correlation functions to show the absence of anomalous canonical commutators for generic Berry's phase.
 
We first rewrite the quadratic action \eqref{action2} in the form
\begin{eqnarray}\label{quadratic action}
S&=&\int dt \frac{1}{2}(x_{k}, p_{k})
\{\left(\begin{array}{cc}
            F_{kl}(\vec{x}_{0})&-\delta_{kl}\\
            \delta_{kl}&-\Omega_{kl}(\vec{p}_{0})
            \end{array}\right) \frac{\partial}{\partial t} 
\nonumber\\
&&-\left(\begin{array}{cc}
            \partial_{k}\partial_{l}\phi(x_{0})&0\\
            0&\frac{1}{m}\delta_{kl}
            \end{array}\right) \}
            \left(\begin{array}{c}
            x_{l}\\
            p_{l}
            \end{array}\right).
\end{eqnarray}
In the sense of the phase space path integral,
we have
the time ordered product as an inverse of the operator appearing in \eqref{quadratic action}
\begin{eqnarray}\label{T product}
&&\int dt e^{i\omega t}
\langle \hat{T}  \left(\begin{array}{c}
            x_{l}(t)\\
            p_{l}(t)
            \end{array}\right)(x_{k}(0), p_{k}(0))\rangle\nonumber\\
&&=i\hbar\{\left(\begin{array}{cc}
            F_{kl}(\vec{x}_{0})&-\delta_{kl}\\
            \delta_{kl}&-\Omega_{kl}(\vec{p}_{0})
            \end{array}\right)(-i\omega) -\left(\begin{array}{cc}
            \partial_{k}\partial_{l}\phi(x_{0})&0\\
            0&\frac{1}{m}\delta_{kl}
            \end{array}\right)\}^{-1}.
\end{eqnarray}
 The BJL analysis applied to \eqref{T product} then gives the commutation relations 
\begin{eqnarray}\label{Duval formula}
[ \left(\begin{array}{c}
            x_{l}(0)\\
            p_{l}(0)
            \end{array}\right), (x_{k}(0), p_{k}(0))]= i\hbar M_{lk}
\end{eqnarray}
where we defined  $M_{jk}$ to be the right-hand side of
\begin{eqnarray}\label{Duval}
\left(\begin{array}{cc}
            F_{kl}(\vec{x}_{0})&-\delta_{kl}\\
            \delta_{kl}&-\Omega_{kl}(\vec{p}_{0})
            \end{array}\right)^{-1} = \frac{\left(\begin{array}{cc}
            \epsilon^{ilk}\Omega_{i}&\delta_{lk}+B_{l}\Omega_{k}\\
            -\delta_{lk}-\Omega_{l}B_{k}&-\epsilon^{ilk}B_{i}
            \end{array}\right)}{1+\vec{B}\cdot\vec{\Omega}}\equiv (M_{lk})
\end{eqnarray}
valid for $F_{kl}(\vec{x}_{0})=\epsilon^{klm}B_{m}$ and $\Omega_{kl}(\vec{p}_{0})=\epsilon^{klm}\Omega_{m}$, which has been evaluated in~\cite{Duval}. 

These commutation relations \eqref{Duval formula} naturally reproduce our results \eqref{commutator} and the (quantum version of) Poisson brackets \eqref{Poisson bracket} in the presence of a {\em genuine Dirac monopole.} This agreement illustrates the connection of the phase space path integral and the generalized classical phase space analysis in \cite{Duval}; in both approaches, one considers an inversion of the operator appearing in the action, but the classical approach is more general by treating a general action while the path integral is defined practically for a quadratic action.

The kinetic term $\epsilon_{n}(p)=\frac{p_{k}p_{k}}{2m}$ and the potential $\phi(x)=\frac{1}{2}\partial_{k}\partial_{l}\phi(x_{0})x_{k}x_{l}$ in the simplified action \eqref{action2} do not contribute to the commutators \eqref{Duval formula}.  This explains the well-known (but mysterious) fact that any potential which does not contain time derivative does not modify the Heisenberg commutation relations; similarly, the term $\epsilon_{n}(p)$ does not modify the commutation relations since coordinates and momenta are symmetric in the present phase space path integral. Technically, both of $\phi(x)$ and $\epsilon_{n}(p)$ terms do not contain the factor $\omega$ in \eqref{T product}. In contrast, the Dirac monopole term contributes to the equal-time commutators because of the factor $\omega$ multiplying the magnetic flux  
$\Omega_{kl}(\vec{p}_{0})i\omega$ in 
\eqref{T product}.
Thus the potential and kinetic energy terms, $\phi(x)$ and $\epsilon_{n}(p)$, are not crucial in analyzing the canonical commutators in the phase space formalism.
 
We next illustrate the use of Berry's phase as a higher order correction to the time ordered correlation
function.
If one rewrites the formulas  \eqref{quadratic action} and \eqref{T product} in Appendix by regarding the Dirac monopole term as an interaction term inside the action, one obtains by setting $F_{kl}=\phi(x)=0$,
\begin{eqnarray}\label{perturbation formula}
\int dt e^{i\omega t}\langle \hat{T} x_{l}(t)x_{k}(0)\rangle=\frac{i\hbar}{m}\frac{\delta_{lk}}{\omega^{2}}+(-\frac{\hbar}{\omega})(\frac{-\omega}{\hbar}\Omega_{lk})(-\frac{\hbar}{\omega})
\end{eqnarray}
where the second term is understood as a perturbative correction to  the free correlation \eqref{Green's function} in the first term.

The corresponding Feynman diagram is shown in Fig.3.
\begin{figure}[htb]
 \begin{center}
   \includegraphics*[width = 8.0cm]{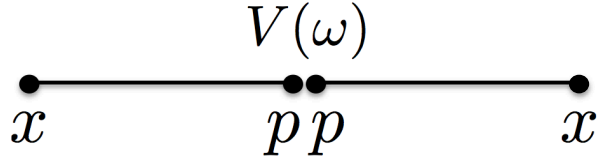}
   \caption{Feynman diagram to evaluate the $O(\hbar)$ quantum correction induced by Berry's phase represented by a vertex $V(\omega)$ to a two-point correlation function 
   $\int dt e^{i\omega t}\langle \hat{T} x_{k}(t)x_{l}(0)\rangle$.  }\label{Fig_03}
 \end{center}
\end{figure}
The free propagator $\langle \hat{T} x_{k}(t)p_{l}(0)\rangle_{0}$ is given in \eqref{Green's function}. The vertex part in momentum representation (Fourier transformed $\omega$-representation) which represents Berry's phase is written as $V(\omega)$ in Fig.3. It is given by 
\begin{eqnarray}\label{monopole-interaction}
V_{kl}(\omega) = \frac{i}{\hbar}(-i\omega)\Omega_{kl}(p_{0})
\end{eqnarray}
in the present case \eqref{perturbation formula}.

The BJL prescription \eqref{BJL2} thus gives
\begin{eqnarray}\label{BJL3}
[x_{k}(0),x_{l}(0)]&=&\lim_{\omega\rightarrow\infty}-i\omega \int dt e^{i\omega t}\langle \hat{T} x_{k}(t)x_{l}(0)\rangle\nonumber\\
&=&\lim_{\omega\rightarrow\infty} i\omega \frac{\hbar}{\omega}\frac{1}{\hbar}\omega\Omega_{kl}(p_{0}) \frac{\hbar}{\omega}\nonumber\\
&=& i\hbar \Omega_{kl}(p_{0})
\end{eqnarray}
which reproduces the order $O(\hbar^{2})$ term in   \eqref{anomalous coordinate commutator} for the case of the genuine Dirac monopole.  

It is useful to analyze the change of canonical commutators when Berry's phase changes its form assumed  at the adiabatic limit to another form at the non-adiabatic limit. 
We thus consider Berry's phase, that assumes a genuine magnetic monopole form in the adiabatic limit but vanishes in the nonadiabatic limit, by introducing a constant positive parameter $\omega_{0}$
\begin{eqnarray}\label{monopole-interaction2}
V_{kl}(\omega) = \frac{1}{\hbar}\omega
\Omega_{kl}(p_{0}) \theta(\omega_{0}-|\omega|),
\end{eqnarray}
which is non-vanishing and assumes the adiabatic form only for $|\omega|\leq \omega_{0}$
instead of \eqref{monopole-interaction}.   One then obtains the vanishing result in \eqref{BJL3}  for any finite value of $\omega_{0}$, namely, we have no anomalous canonical commutators induced by such Berry's phase and thus no deformation of the principle of quantum mechanics. {\em The anomalous commutator \eqref{anomalous coordinate commutator} is realized only when Berry's phase is assumed to have the precise Dirac monopole form even in the non-adiabatic limit, which corresponds to $\omega_{0}=\infty$}. We emphasize that 
one is assuming this property of Berry's phase implicitly when one writes the effective action \eqref{action} with a genuine Dirac monopole. The actual value of $\omega_{0}$ which defines the validity domain of the adiabatic approximation is small; for example, the ideal adiabatic limit corresponds to $\omega_{0} \rightarrow  0$ or $T_{0}=2\pi/\omega_{0} \rightarrow \infty$ \cite{Simon}. This analysis shows that the anomalous commutator is determined by  the behavior of  Berry's phase at the  non-adiabatic (high frequency ) limit. 
If the quantization should {\em not} be determined by the  large $\omega$ behavior, contrary to the BJL prescription, the universal validity of quantum commutators under various deformations of the Hamlitonian such as the change of (time independent) potentials would not hold.

The generic property of Berry's phase analyzed in Sections 1 and 2 is that Berry's phase approaches the exact monopole form at the adiabatic limit but it generally vanishes at the non-adiabatic limit. The behavior of \eqref{monopole-interaction2} with small $|\omega|\leq \omega_{0}$ is thus universal for Berry's phase in the adiabatic limit. The vanishing of  \eqref{monopole-interaction2} at large $\omega$ will simulate the vanishing Berry's phase in the non-adiabatic limit as in \eqref{trivial phase} and \eqref{trivial phase2}. The present perturbative treatment of the induced Berry's phase also shows that generic Berry's phase does not lead to anomalous canonical commutators \eqref{anomalous coordinate commutator}.


\begin{thebibliography}{99}
\bibitem{Higgins}
H. Longuet-Higgins, Proc. Roy. Soc. A{\bf 344}, 147 (1975).
\bibitem{Berry}
M. V.  Berry, Proc. R. Soc. Ser. A{\bf 392}, 45 (1984) .
\bibitem{Simon}
B. Simon, Phys. Rev. Lett. {\bf 51}, 2167 (1983).
\bibitem{fujikawa-ap2007}
K. Fujikawa, Ann. of Phys. {\bf 322}, 1500 (2007).
\bibitem{Deguchi-Fujikawa-2019}
S. Deguchi and K. Fujikawa, Phys. Rev. D{\bf 100},  025002 (2019).
\bibitem{Pancharatnam}
S. Pancharatnam, Proc. Indian Acad. Sci. A{\bf 44}, 247 (1956).
\bibitem{Dirac}
P.A.M. Dirac, Proc. Roy. Soc.  Landon {\bf 133}, 60 (1931).
\bibitem{Karplus}
T. Jungwirth, Q. Niu, A. H. MacDonald, Phys. Rev. Lett. {\bf 88}, 207208 (2002).
\bibitem{Fang}
Z. Fang, et al., Science {\bf 302}, 92 (2003), and references therein.
\bibitem{Hirsch}
J. E. Hirsch, Phys. Rev. Lett. {\bf 83}, 1834 (1999).\\
S.-F. Zhang, Phys. Rev. Lett. {\bf 85}, 393 (2000).\\
S. Murakami, N. Nagaosa, and S.-C. Zhang, Science {\bf 301}, 1348 (2003).
\bibitem{Stone}
M. Stone, Phys. Rev. {\bf D 33}, 1191 (1986).
\bibitem{deguchi}
K. Fujikawa, Mod. Phys. Lett. A{\bf 20},  335 (2005).\\
S. Deguchi and K. Fujikawa, Phys. Rev. A{\bf 72}, 012111 (2005). \\
K. Fujikawa,   Phys. Rev. D{\bf 72},  025009 (2005).
\bibitem{Niu}
D. Xiao, J. Shi, and Q. Niu, Phys. Rev. Lett. {\bf 95}, 137204
(2005).
\bibitem{Wu-Yang}
T. T.  Wu and C. N. Yang, Phys. Rev. D{\bf 12}, 3845 (1975).
\bibitem{Duval}
C. Duval, Z. Horvath, P. A. Horvathy, L. Martina, and
P. Stichel, Mod. Phys. Lett. B{\bf 20}, 373 (2006); Phys. Rev. Lett. {\bf 96}, 099701 (2006).
\bibitem{Faddeev-Jackiw}
L. D. Faddeev and R. Jackiw, Phys. Rev. Lett. {\bf 60}, 1692 (1988).
\bibitem{BJL}
J. Bjorken, Phys. Rev. {\bf 148},  1467 (1966).\\
K. Johnson and F. Low, Prog. Theor. Phys. Suppl. {\bf 37-38}, 74 (1966).
\bibitem{fujikawa-prd2018}
K. Fujikawa, Phys. Rev. D{\bf 97},  016018  (2018).
\bibitem{Baer}
M. Baer and G. D. Billing, Special issue on The Role of Degenerate States in Chemistry, Advances in Chemical Physics, Vol. {\bf 124} (Wiley, Hoboken, New Jersey, USA, 2002).
\bibitem{Nambu}
Y. Nambu, Prog. Theor. Phys. {\bf 7}, 131 (1952).\\
K. Fujikawa, Phys. Rev. D{\bf 70}, 085006 (2004).
\bibitem{Jackiw}
 R. Jackiw and K. Johnson, Phys. Rev. {\bf 182}, 1459 (1969).
\bibitem{Adler}
S. Adler, in `` Lecture on Elementary Particles and
Quantum Field Theory'', eds. S. Deser et al. (MIT Press, Mass., 1970).
\bibitem{FS}
K. Fujikawa and H. Suzuki, "Path Integrals and Quantum Anomalies", (Oxford University Press, 2004).

\end{thebibliography}
\end{document}